\documentclass[
  aps,
  prl,
  showpacs,
  superscriptaddress,
  floatfix,
  twocolumn,
  secnumarabic,
  amssymb,
  amsmath]{revtex4-1}

\usepackage{graphicx}
\usepackage{bm}
\usepackage{color,ulem}
\usepackage{lineno}
\usepackage{upgreek}
\usepackage[colorlinks=true,citecolor=blue,linkcolor=blue,pdfstartview=FitH,pdfauthor={Klaeui, Goennenwein, Gepraegs}]{hyperref}

\begin{document}

\title{Origin of the spin Seebeck effect probed by temperature dependent measurements in Gd$_{3}$Fe$_{5}$O$_{12}$}

\author{Stephan~Gepr{\"a}gs}
\thanks{These authors contributed equally to the work.}
\affiliation{Walther-Mei{\ss}ner-Institut, Bayerische Akademie der Wissenschaften, 85748 Garching, Germany}
\author{Andreas~Kehlberger}
\altaffiliation[]{Graduate School Materials Science in Mainz, Staudinger Weg 9, 55128, Germany}
\thanks{These authors contributed equally to the work.}
\affiliation{Institute of Physics, University of Mainz, 55099 Mainz, Germany}
\affiliation{Materials Science in Mainz, Staudinger Weg 9, 55128 Mainz, Germany}
\author{Tomek~Schulz}
\affiliation{Institute of Physics, University of Mainz, 55099 Mainz, Germany}
\author{Christian~Mix} 
\affiliation{Institute of Physics, University of Mainz, 55099 Mainz, Germany}
\affiliation{Materials Science in Mainz, Staudinger Weg 9, 55128 Mainz, Germany}
\author{Francesco~Della~Coletta}
\affiliation{Walther-Mei{\ss}ner-Institut, Bayerische Akademie der Wissenschaften, 85748 Garching, Germany}
\author{Sibylle~Meyer}
\affiliation{Walther-Mei{\ss}ner-Institut, Bayerische Akademie der Wissenschaften, 85748 Garching, Germany}
\author{Akashdeep~Kamra}
\affiliation{Walther-Mei{\ss}ner-Institut, Bayerische Akademie der Wissenschaften, 85748 Garching, Germany}
\affiliation{Kavli Institute of Nanoscience, Delft University of Technology, Delft, The Netherlands}
\author{Matthias~Althammer}
\affiliation{Walther-Mei{\ss}ner-Institut, Bayerische Akademie der Wissenschaften, 85748 Garching, Germany}
\author{Gerhard~Jakob}
\affiliation{Institute of Physics, University of Mainz, 55099 Mainz, Germany}
\affiliation{Materials Science in Mainz, Staudinger Weg 9, 55128 Mainz, Germany}
\author{Hans Huebl}
\affiliation{Walther-Mei{\ss}ner-Institut, Bayerische Akademie der Wissenschaften, 85748 Garching, Germany}
\affiliation{Nanosystems Initiative Munich (NIM), Schellingstra{\ss}e 4, 80799 M\"unchen, Germany}
\author{Rudolf~Gross}
\affiliation{Walther-Mei{\ss}ner-Institut, Bayerische Akademie der Wissenschaften, 85748 Garching, Germany}
\affiliation{Nanosystems Initiative Munich (NIM), Schellingstra{\ss}e 4, 80799 M\"unchen, Germany}
\affiliation{Physik-Department, Technische Universit\"at M\"unchen, 85748 Garching, Germany}
\author{Sebastian~T.B.~Goennenwein}
\affiliation{Walther-Mei{\ss}ner-Institut, Bayerische Akademie der Wissenschaften, 85748 Garching, Germany}
\affiliation{Nanosystems Initiative Munich (NIM), Schellingstra{\ss}e 4, 80799 M\"unchen, Germany}
\author{Mathias~Kl\"aui}
\thanks{Corresponding authors: Stephan.Gepraegs@wmi.badw.de, Sebastian.Goennenwein@wmi.badw-muenchen.de, and Klaeui@uni-mainz.de}
\affiliation{Institute of Physics, University of Mainz, 55099 Mainz, Germany}
\affiliation{Materials Science in Mainz, Staudinger Weg 9, 55128 Mainz, Germany}

\date{\today}

\begin{abstract}
We probe the spin Seebeck effect in Gd$_{3}$Fe$_{5}$O$_{12}$/Pt hybrid structures as a function of temperature and observe two sign changes of the spin Seebeck signal with decreasing temperature. A first sign change occurs at a temperature close to the Gd$_{3}$Fe$_{5}$O$_{12}$ magnetic compensation point at around 280\,K. There the spin Seebeck signal changes sign abruptly with unaltered amplitude, indicating that the spin current is mainly caused by the magnetic Fe sub-lattices, which reorient their directions at this temperature. A second, more gradual sign change takes place around the ordering temperature of the Gd sub-lattice in the range of 65-85\,K, showing that the Gd magnetic sub-lattice dominates the thermally driven spin current at lower temperatures. These sign changes together with the non-monotonous dependence of the spin Seebeck signal on the temperature demonstrate that the magnonic spin current is not simply replicating the effective magnetization of Gd$_{3}$Fe$_{5}$O$_{12}$. Rather, the thermally generated net spin current results from a complex interplay of the three magnetic sub-lattices involved.  
\end{abstract}


\maketitle

Magnons are the fundamental excitations in a magnetically ordered system. Their dispersion relations, propagation properties, etc., have been extensively investigated theoretically and experimentally.\cite{Cherepanov:PhysRep:1993,Serga:JPhysD:2010,Kruglyak:2010:JPhysD} In the presence of a temperature gradient, the magnon population at the hot end of a magnetic material is larger than at the cold end, resulting in an effective magnon flow along the thermal gradient.\cite{Uchida:NatureMater:2010,Uchida:APL:2010,Hinzke:PRL:2011,Bauer:NatureMater:2012,Ritzmann:PRB:2014} Since magnons carry spin angular momentum, a thermal gradient thus will drive a (pure) magnonic spin current in a magnetic material, a phenomenon referred to as the spin Seebeck effect (SSE).\cite{Uchida:NatureMater:2010,Bauer:NatureMater:2012}

\begin{figure}[tbh]
  \includegraphics[width=1.0\columnwidth]{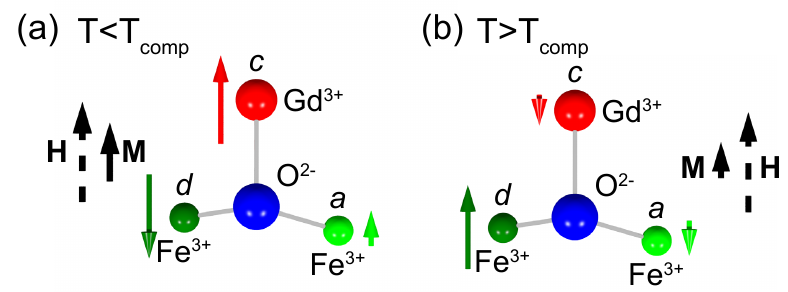}
    \caption{(color online) (a), (b) The three different magnetic sub-lattices of GdIG shown for temperatures below ($T<T_\mathrm{comp}$) and above ($T>T_\mathrm{comp}$) the magnetic compensation temperature $T_\mathrm{comp}.$\cite{Dionne:book:2009} Due to the strong temperature dependence of the magnetic Gd sub-lattice ($c$ sites), the Gd sub-lattice dominates the magnetic behavior for $T<T_\mathrm{comp}$, while for $T>T_\mathrm{comp}$ the magnetic behavior is mainly caused by the antiferromagnetically coupled Fe sub-lattices ($d$ and $a$ sites). In the presence of a finite magnetic field, the net magnetization $\mathbf{M}$ points along the external magnetic field $\mathbf{H}$. The direction and length of the colored arrows mark the direction and magnitude of the magnetization for the three magnetic sub-lattices.}
    \label{fig:Fig1}
\end{figure}

Lately, the SSE has attracted much attention, due to open fundamental physics questions regarding its origin on the one hand,\cite{Uchida:APL:2010,Ohnuma:PRB:2013,Kehlberger:arXiv:2013,Schreier:PRB:2013,Siegel:SciRep:2014,Xiao:PRB:2010,Sandweg:PRL:2011,Tikhonov:NatureComm:2013,Agrawal:PRL:2013,Agrawal:arXiv:2013,Roschewsky:arXiv:2013} and due to its application potential on the other hand.\cite{Kirihara:NatureMater:2012} It is now widely accepted that the thermopower signals measured in magnetic insulator/normal metal hybrids in the so-called longitudinal SSE geometry are indeed a consequence of the magnonic spin currents generated by the temperature gradient.\cite{Uchida:APL:2010,Ohnuma:PRB:2013,Kehlberger:arXiv:2013,Schreier:PRB:2013,Siegel:SciRep:2014} However, the exact mechanism behind the excitation of the individual magnon modes, and the contribution of different magnon modes to the spin current are still under discussion.\cite{Xiao:PRB:2010,Sandweg:PRL:2011,Tikhonov:NatureComm:2013,Agrawal:PRL:2013,Agrawal:arXiv:2013,Roschewsky:arXiv:2013} So far most SSE experiments were performed in electrically insulating garnet materials, which are ferrimagnetic, with several interacting magnetic sub-lattices.\cite{Pauthenet:JAP:1958,Winkler:book:1981,Smit:book:1958,Cherepanov:PhysRep:1993} For yttrium iron garnet (Y$_{3}$Fe$_{5}$O$_{12}$, YIG) it was shown that the two Fe sub-lattices with 20 magnetic atoms per unit cell lead to a complex spin wave dispersion relation.\cite{Cherepanov:PhysRep:1993} Even more complex magnon spectra are expected for iron garnets $RE_{3}$Fe$_{5}$O$_{12}$ where the rare earth $RE$ carries a magnetic moment (e.g., $RE=$Gd, Tb, Dy, etc.), such that one can anticipate more complex and rich SSE physics in these materials (cf.~Fig.~\ref{fig:Fig1}). In particular, Ohnuma \textit{et al.} theoretically calculated the SSE voltage amplitude and sign depending on the magnetization in a ferrimagnet.\cite{Ohnuma:PRB:2013} Using a two-sub-lattice model, these authors predict that the SSE signal should reverse sign at the magnetization compensation point, due to the different, non-degenerate magnons that contribute to the spin current. Since real garnets exhibit a rather complex magnon dispersion relation with a large number of magnon branches,\cite{Cherepanov:PhysRep:1993} however, such a two-sub-lattice model might not be sufficient to fully describe the real material. Moreover, the detection of the magnonic spin currents is usually realized via the inverse spin Hall effect (ISHE) in an adjacent normal metal, and the spin current transfer (spin mixing conductance) of magnons carried by different magnetic sub-lattices or atomic species across the interface, from the magnetic insulator to the normal metal, might also be different. Thus, there is a clear need to study the SSE in complex ferrimagnetic garnets and magnetic systems with complex magnetic structure.

In this paper, we experimentally investigate the evolution of the magneto-thermopower (i.e., the longitudinal SSE) as a function of temperature in Gd$_3$Fe$_5$O$_{12}$/Pt (GdIG/Pt) hybrids. We observe a rather abrupt SSE sign change close to the magnetic compensation point, followed by yet another, more gradual SSE sign change at lower temperatures. We interpret our experimental observations in terms of the different relative contributions of the Fe and Gd magnons to the SSE, which have qualitatively different temperature dependences. 

\begin{figure}[tbh]
  \includegraphics[width=1.0\columnwidth]{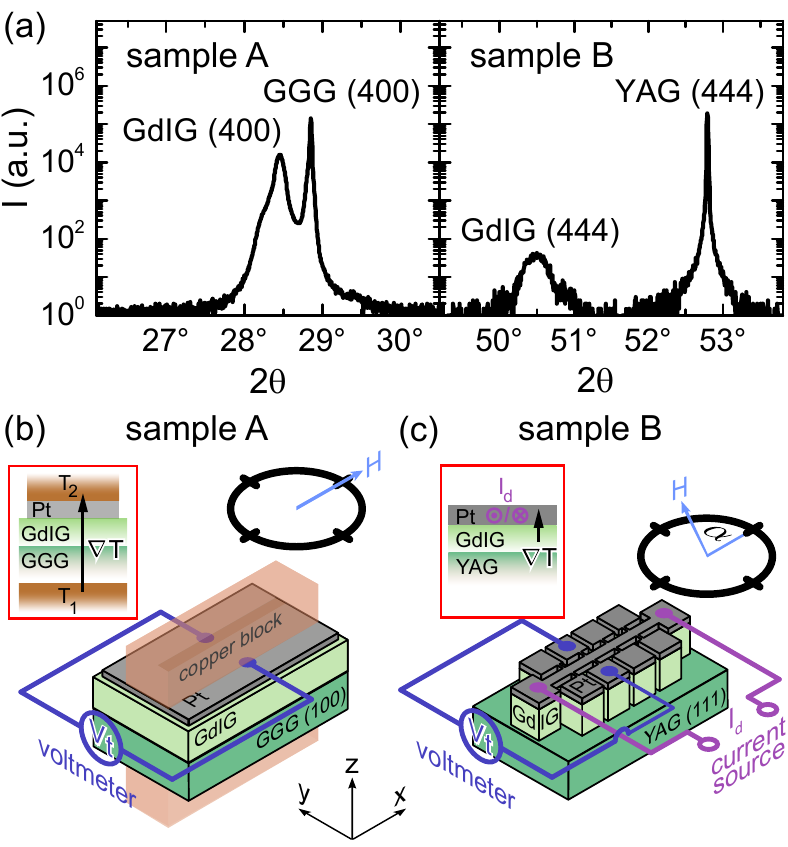}
    \caption{(color online) (a) High resolution x-ray diffraction (HR-XRD) measurements of sample A and sample B. (b)-(c) Sketches of the experimental setups used to record the longitudinal spin Seebeck effect (SSE).}
    \label{fig:Fig2}
\end{figure}

Two sets of samples were fabricated and investigated. At Mainz, GdIG films were deposited on single crystalline, $[100]$-oriented Gadolinium Gallium Garnet (Gd$_{3}$Ga$_{5}$O$_{12}$, GGG) substrates by pulsed laser deposition (PLD) using a KrF excimer laser with a laser fluence of $1.2\,\mathrm{J}/\mathrm{cm}^2$. We here focus on a $100\,\mathrm{nm}$ thick GdIG film (sample A) grown with optimized parameters,~i.e., in an oxygen atmosphere at $2\times 10^{-2}\,\mathrm{mbar}$ and at a substrate temperature of $650^\circ\mathrm{C}$. No parasitic phases were detected using x-ray diffraction [cf.~Fig.~\ref{fig:Fig2}(a)]. Furthermore, an out-of-plane lattice constant of $(1.254\pm 0.005)\,\mathrm{nm}$, which is comparable to the bulk value of $1.247\,\mathrm{nm}$,\cite{Geller:PR:1961} is calculated using the GdIG $(400)$ film reflection. Moreover, the rocking curve around the GdIG~(400) reflection has a full width at half maximum (FWHM) of $0.036^\circ$ indicating a high crystalline quality and low mosaicity. The $8\,\mathrm{nm}$ thick Pt layer, used for the ISHE measurements, was deposited by DC-magnetron sputtering after \textit{in-situ} cleaning of the interface. At Garching, the GdIG films were deposited onto single crystalline, $[111]$-oriented Yttrium Aluminum Garnet (Y$_{3}$Al$_{5}$O$_{12}$, YAG) substrates via PLD. Best structural and magnetic properties of the GdIG films were obtained by using a substrate temperature of $500^\circ\mathrm{C}$, an oxygen atmosphere of $1.25\times 10^{-2}\,\mathrm{mbar}$, and an energy fluence of the KrF excimer laser of $2.0\,\mathrm{J}/\mathrm{cm}^2$ at the target surface.\cite{Gepraegs:APL:2012} The crystalline quality of the GdIG films demonstrated by the FWHM of the rocking curves around the GdIG (444) reflection of around $0.04^\circ$ is comparable to the GdIG thin films fabricated at Mainz. We here discuss a $26\,\mathrm{nm}$ thick GdIG film with a lattice constant of $(1.252 \pm 0.005)\,\mathrm{nm}$ (sample B), covered \textit{in-situ}, without breaking the vacuum, with $4\,\mathrm{nm}$ of Pt deposited via electron beam evaporation [cf.~Fig.~\ref{fig:Fig2}(a)].

The longitudinal SSE was recorded as a function of temperature, using two complementary methods. At Mainz, the sample is sandwiched between two copper blocks,\cite{Kehlberger:arXiv:2013} which are cooled by the Helium exchange gas in a magnet cryostat [cf.~Fig.~\ref{fig:Fig2}(b)]. A resistive heater within each block allows to independently set and stabilize its temperature. In particular, it thus is possible to invert the temperature difference between the two copper blocks, allowing to exclude spurious effects due to asymmetric sample mounting. The temperature of the copper blocks is controlled by Pt-100 elements, which are also used to define the temperature difference applied across the sample. Thin single crystalline sapphire plates combined with thermo-conductive foil ensure good thermal contact as well as electric insulation between the copper blocks and the sample. To determine the longitudinal SSE signal, the transverse voltage $V_{\mathrm{t}}$, perpendicular to an external magnetic field, is measured while modifying the magnetic field magnitude at a fixed magnetic field orientation ($\mathbf{H}\parallel \mathbf{x}$). In Garching, the GdIG/Pt bilayer is micropatterned into a Hall bar structure using optical lithography and Argon ion beam milling [cf.~Fig.~\ref{fig:Fig2}(c)]. The sample is then inserted into a 3D vector magnet cryostat, in which He exchange gas in a variable temperature insert allows us to adjust the sample base temperature. We generate the temperature gradient across the GdIG/Pt interface required for SSE experiments by sourcing a large current $I_{\mathrm{d}}$ along the Pt microstructure itself, and exploit the temperature-dependent resistance of the Pt for on-chip thermometry.\cite{Schreier:APL:2013} Here, $V_{\mathrm{t}}$ is recorded as a function of the in-plane external magnetic field direction $\alpha$ using a fixed magnetic field magnitude. As detailed below, both experimental methods yield qualitatively and quantitatively similar results, demonstrating the validity of the approaches used.

\begin{figure}[thb]
  \includegraphics[width=0.90\columnwidth]{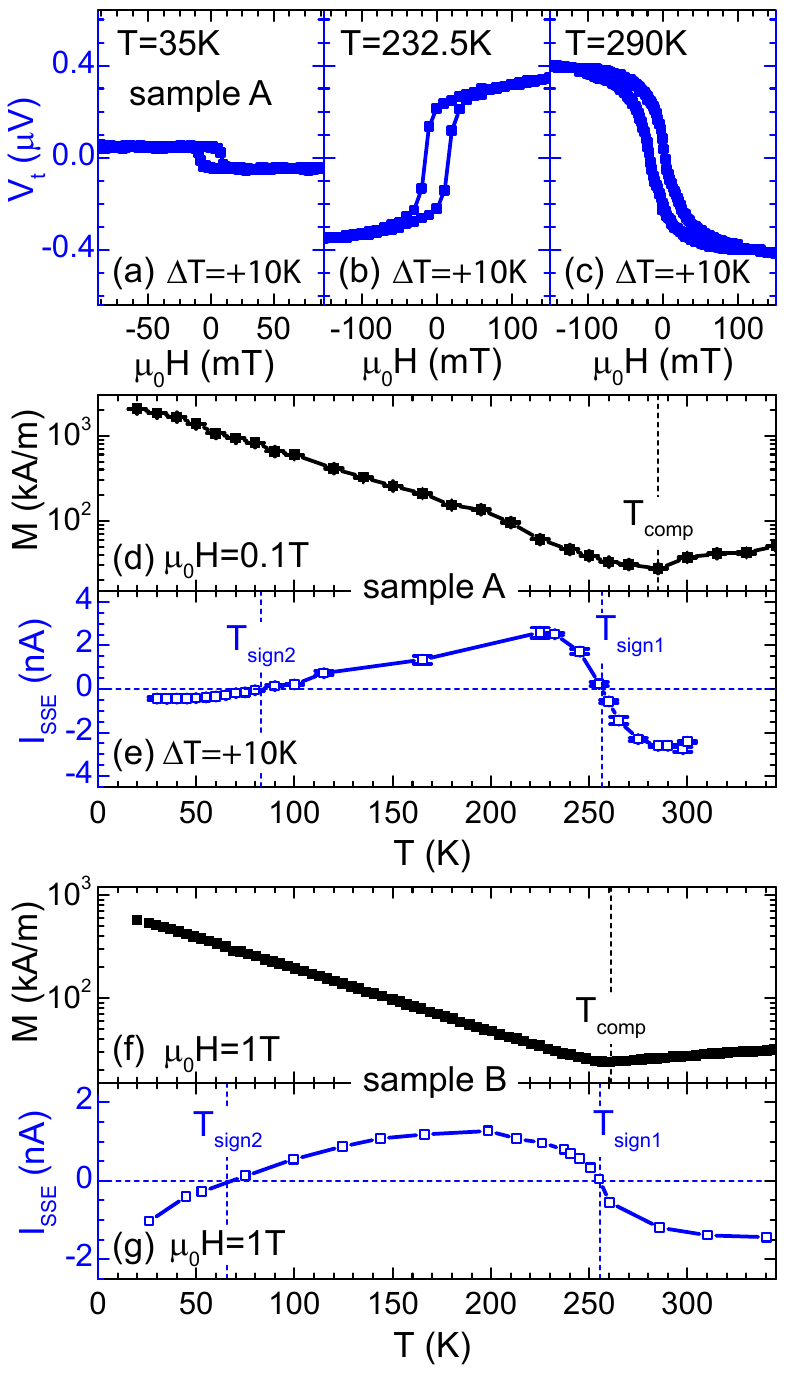}
    \caption{(color online) (a)-(c) Measured transverse voltage $V_\mathrm{t}$ of sample~A for selected temperatures showing two sign changes as a function of temperature. (d) Temperature-dependent magnetization of sample~A recorded at a magnetic field of 0.1\,T. (e) SSE signal $I_\mathrm{SSE}$ as a function of temperature obtained from the difference in $V_\mathrm{t}$ at positive and negative saturation taking the temperature dependence of the Pt resistance $R(T)$ into account. (f) Magnetization as a function of temperature of sample~B measured at $\mu_0 H = 1$\,T. (g) $I_\mathrm{SSE}$ signal of sample B derived from angular dependent measurements. Here, the $V_\mathrm{t}$ is recorded under different in-plane orientations $\alpha$ of the magnetic field at constant magnitude of 1\,T. The SSE signal $I_\mathrm{SSE}=V_\mathrm{SSE}/R(T)$ is calculated from $V_\mathrm{SSE}=\frac{1}{2} \left[ V_\mathrm{t}(+I_\mathrm{d})+V_\mathrm{t}(-I_\mathrm{d}) \right]$. The blue dashed lines mark the zero-crossings of the $I_\mathrm{SSE}$ signal. The temperature of the magnetic compensation points $T_\mathrm{comp}$ for sample A and B are indicated by the black dashed lines.}
    \label{fig:Fig3}
\end{figure}

First, we investigate the magnetic properties of the GdIG films as a function of temperature, recorded using SQUID magnetometry. As evident from Figs.~\ref{fig:Fig3}(d) and (f), sample A and sample B show the characteristics expected for a ferrimagnet with a compensation point:\cite{Dionne:book:2009} The magnetization first decreases with decreasing temperature, reaches  a minimum, and then increases again. As sketched in Figs.~\ref{fig:Fig1}(a) and (b), GdIG is a ferrimagnet with three magnetic sub-lattices. Each unit cell consists of 12 trivalent Fe atoms that are tetrahedrally coordinated with oxygen atoms ($d$ sites, point group $S_4$), 8 Fe atoms that are octahedrally coordinated ($a$ sites, point group $S_6$), and 12 dodecahedrally coordinated Gd atoms ($c$ sites). The two Fe sub-lattices are strongly coupled via antiferromagnetic superexchange (exchange constant $J_{ad} \simeq 32$\,cm$^{-1}$),\cite{Harris:PR:1963} with a N\'{e}el temperature of about $T_{\mathrm{N}}=550\,\mathrm{K}$. Since the Gd moments are more weakly exchange coupled to the Fe $a$ sub-lattices ($J_{ac} \simeq 7$\,cm$^{-1})$\cite{Harris:PR:1963} leading to an ordering temperature of the Gd spins at around 69\,K,\cite{Goulon:NewJPhys:2012} they experience an effective magnetic field below $T_{\mathrm{N}}$ and can be treated as an `exchange-enhanced' paramagnetic moment.\cite{Belov:PhysUsp:1996} As a consequence the Gd magnetization can be described by a strongly temperature-dependent Brillouin function for spin 7/2. At high temperatures, the $d$ site Fe ions dominate the net magnetization of GdIG. As the temperature is reduced, the magnetization of the Gd sub-lattice strongly increases, and together with the Fe magnetization at the $a$ site eventually overwhelms the Fe magnetization at the $d$ site. At the so-called magnetic compensation temperature $T_{\mathrm{comp}}\approx 288\,\mathrm{K}$ (bulk value),\cite{Pauthenet:JAP:1958,Rodic:SolidStateComm:1990} the magnetization of the $a$ site of Fe and the $c$ site of Gd is equal in magnitude but antiparallel to the magnetization of the $d$ site of Fe (cf.~Fig.~\ref{fig:Fig1}), such that the remanent magnetization of GdIG becomes zero. Since our magnetometry experiments were performed in a finite external magnetic field of magnitude 0.1-1\,T (comparable to the experimental situation in the SSE experiments below), the net GdIG magnetization does not vanish, but rather goes through a minimum at $T_{\mathrm{comp}}$, owing to the paramagnetic response of the Gd moments. Figure~\ref{fig:Fig3}(d) reveals that the compensation temperature of sample A is around $285\,\mathrm{K}$, in line with literature.\cite{Rodic:SolidStateComm:1990,Pauthenet:JAP:1958} For sample B, $T_{\mathrm{comp}}\approx 260\,\mathrm{K}$ is somewhat lower [cf.~Fig.~\ref{fig:Fig3}(f)], most likely due to the smaller GdIG film thickness resulting in a finite strain state, which reduces the magnetic exchange strength. 

We now turn to the SSE experiments shown in Figs.~\ref{fig:Fig3}(a)-(c). The data was taken on sample A at four different sample base temperatures, with a temperature difference of $+10\,\mathrm{K}$ applied across the sample. Positive temperature difference hereby means that the Pt is warmer than the GdIG and the GGG substrate. As a function of the external magnetic field, the voltage signal $V_{\mathrm{t}}$ shows a hysteresis behavior, which is  characteristic for the SSE.\cite{Uchida:APL:2010} Interestingly, however, the $V_{\mathrm{t}}$ hysteresis loop flips sign two times with decreasing temperature: while $V_{\mathrm{t}}(H)$ is `regular' for $T=280\,\mathrm{K}$ (i.e., the $V_{\mathrm{t}}(H)$ recorded in GdIG/Pt hybrids has the same sign as the established SSE signal of YIG/Pt),\cite{Schreier:arXiv:2014} the $V_{\mathrm{t}}(H)$ loop recorded for $T=232.5\,\mathrm{K}$ is `inverted', while the loop at $T=35\,\mathrm{K}$ again is `regular'.

In order to characterize the temperature-dependent evolution of the SSE signal in more detail, we define the SSE amplitude $V_{\mathrm{SSE}}$ as the difference $V_{\mathrm{SSE}}=(V_{\mathrm{t}}(+H_{\mathrm{sat}})-V_{\mathrm{t}}(-H_{\mathrm{sat}}))/2$, recorded at external magnetic field strengths $\mu_0 H_{\mathrm{sat}}$ large enough to saturate the GdIG magnetization. Taking the temperature dependence of the Pt resistance $R(T)$ into account, Figure~ \ref{fig:Fig3}(e) shows the $I_{\mathrm{SSE}}(T)=V_{\mathrm{SSE}}(T)/R(T)$ curve from sample A. The SSE signal first is negative at high temperatures, then becomes positive in a temperature interval $T_{\mathrm{sign2}} \lesssim T \lesssim T_{\mathrm{sign1}}$ with $T_{\mathrm{sign2}}\approx 80\,\mathrm{K}$, and is negative again for $T\lesssim T_{\mathrm{sign2}}$. Note also that the sign change around $T_{\mathrm{sign1}}\approx 256.5\,\mathrm{K}$ is rather abrupt, while the second sign change around $T_{\mathrm{sign2}}$ is much smoother. As evident from Figure~ \ref{fig:Fig3}(g) the SSE signal $I_{\mathrm{SSE}}$ of sample B derived from angular dependent measurements confirms the described temperature dependence of sample A. The $I_{\mathrm{SSE}}$ signal was obtained by recording the transverse voltage $V_\mathrm{t}$ with $I_\mathrm{d}=6$\,mA applied across the Hall bar structure as a function of the in-plane orientation $\alpha$ of the external magnetic field at constant magnitude of 1\,T. The SSE signal $I_\mathrm{SSE}=V_{\mathrm{SSE}}(T)/R(T)$ was then calculated from $V_\mathrm{SSE}=\frac{1}{2} \left[ V_\mathrm{t}(+I_\mathrm{d})+V_\mathrm{t}(-I_\mathrm{d}) \right]$.\cite{Schreier:APL:2013}

We start the discussion by comparing our experimental observations with the theoretical model by Ohnuma \textit{et al.}.\cite{Ohnuma:PRB:2013} For the compensated ferrimagnet $\mathrm{Er}_{3}\mathrm{Fe}_{5}\mathrm{O}_{12}$, these authors predict a temperature-independent SSE magnitude, which abruptly changes sign once at the magnetization compensation temperature. Ohnuma \textit{et al.} attribute this SSE sign change to the inversion of the sub-lattice magnetization orientations at the compensation point [cf.~Figs.~\ref{fig:Fig1}(a),(b)] corresponding to a reversal of the spin current polarization, while the contributions of the two non-degenerate magnon branches do not change as a function of temperature. This model is based on two non-degenerate magnon branches corresponding to two magnetic sub-lattices, namely a ferric and a rare-earth sub-lattice.\cite{Ohnuma:PRB:2013,Chikazumi:book:1997} While the model by Ohnuma \textit{et al.} thus accounts for the rather abrupt sign change in $I_{\mathrm{SSE}}$ around $T_{\mathrm{sign1}}$, it fails to reproduce the second SSE sign change at $T_{\mathrm{sign2}}$.

Based on these findings, we extend the theory of Ohnuma \textit{et al.}, using a mean field description of the three magnetic sub-lattices in GdIG as shown in Fig.~\ref{fig:Fig1} (cf.~Ref.~\onlinecite{Dionne:book:2009}). We assume that all thermal magnetic excitations in the different magnetic sub-lattices contribute equally to the SSE -- however only if the corresponding sub-lattice is sufficiently ordered. In other words, the magnons in the Fe sub-lattices will substantially contribute to the SSE for $T <  T_{\mathrm{C}} \approx 550\,\mathrm{K}$, while the Gd sub-lattice exhibits a qualitatively different thermomagnetic behavior and significantly contributes to the SSE only for $T\lesssim T_{\mathrm{Gd}} \approx 65-85\,\mathrm{K}$. As mentioned above, the latter temperature is often also referred to as the Gd ordering temperature in the literature.\cite{Goulon:NewJPhys:2012} This assumption is motivated by the existing literature investigating finite temperature effects on ferromagnets via the self consistent renormalization of magnon dispersions.\cite{Bloch:PRL:1962} In particular, it has been shown that the magnons cease to be the admissible eigenstates above a temperature around the ordering temperature. In this scenario, only the magnons from the Fe sub-lattices contribute to the SSE for $T_{\mathrm{Gd}} < T \lesssim T_{\mathrm{Fe}}$. The net Fe magnetization thereby determines the spin current spin polarization, such that the inversion of the sub-lattice magnetization orientation at the magnetization compensation point results in a sharp inversion of the SSE sign at this temperature. For $T \lesssim T_{\mathrm{Gd}}$, however, the Gd magnons become gradually more important. Since the Gd sub-lattice magnetization is oriented along the external magnetic field at these temperatures, the corresponding spin current is carrying angular momentum with an opposite direction compared to the spin current from the Fe magnons. With decreasing $T$, the Gd magnons eventually dominate, resulting in a second, more gradual SSE sign change back to the high temperature sign of the signal, as observed in experiment. This means that the difference in abruptness of the transitions at $T_{\mathrm{sign1}}$ and $T_{\mathrm{sign2}}$ directly reflects the different nature of the transitions ($T_{\mathrm{sign1}}$ related to the abrupt Fe sublattice reorientation at the magnetic compensation point while $T_{\mathrm{sign2}}$ comes from the gradually increase of the influence of the magnetic Gd sub-lattice). Note also that the Gd is weakly exchange-coupled to the Fe sub-lattice, such that magnetic excitations caused by the Gd sub-lattice could arise already for $T>T_{\mathrm{Gd}}$.

Clearly, the above model is too simplistic to quantitatively describe the observed behavior. For a consistent and robust analysis, the full spin wave spectrum needs to be calculated by taking into account all 32 magnetic atoms in the unit cell (8 Fe atoms in the octahedral sites, 12 Fe atoms in the tetrahedral sites and 12 Gd atoms), which is beyond the scope of this paper. Nevertheless, our results clearly show that the magnon spectrum and the temperature dependences of the individual magnon branches are more complex than assumed in Ref.~\onlinecite{Ohnuma:PRB:2013}, which can also contribute to the observed fact that the transition temperature $T_{\mathrm{sign1}}$ is lower than the compensation temperature $T_{\mathrm{comp}}$. Finally we note that a quantitative analysis might also need to consider possibly different spin mixing conductances for the interface between Gd and Pt as compared to Fe and Pt.\cite{Weiler:PRL:2013} 

In summary, we have experimentally investigated the temperature dependence of the longitudinal spin Seebeck effect in GdIG/Pt hybrids with different crystallographic orientations of the GdIG thin films grown on different substrate materials. Using two complementary experimental approaches to record the SSE, two consecutive sign changes of the SSE signal were found with decreasing temperature, which we attribute to different temperature dependences of magnonic spin currents of the Fe and Gd magnetic sub-lattices involved. The first sign change occurs around the magnetic compensation point of GdIG indicating the dominant role of the Fe sub-lattices in this temperature range. The more gradual sign change of the SSE signal at around the ordering temperature of Gd (65-85\,K) demonstrates that the magnetic Gd sub-lattice dominates the spin current at lower temperatures. Our results thus show that magnons from different magnetic sub-lattices can quantitatively alter the SSE and that the thermal magnonic spin currents do not simply replicate the total magnetization. This opens intriguing perspectives for spin caloritronics in ferri- and antiferromagnets and shows that current theoretical models are not sufficient to describe the experimental observations, calling for more elaborate theoretical work.

The authors would like to thank the Deutsche Forschungsgemeinschaft (DFG) for financial support via SPP 1538 	``Spin Caloric Transport'' (Project No. GO 944/4-1 and KL 1811/7-1) and the Graduate School of Excellence Materials Science in Mainz (MAINZ) GSC 266, the EU (IFOX, NMP3-LA-2012 246102; MASPIC, ERC-2007-StG 208162; InSpin FP7-ICT-2013-X 612759) and the Research Center of Innovative and Emerging Materials. At Garching, we thank T.~Brenninger for technical support, A.~Habel and K.~Helm--Knapp for the fabrication of the stoichiometric GdIG target, and M.~Schreier as well as J.~Lotze for fruitful discussions. 


%

\end{document}